\documentclass[runningheads,envcountsame,envcountsect,oribibl]{llncs}

\usepackage[latin1]{inputenc}
\usepackage[dvips]{graphics}
\usepackage{psfrag}
\usepackage{amsmath}
\usepackage{amssymb}
\usepackage[english]{babel}
\usepackage{url}
\usepackage[colorlinks=true,ps2pdf=true]{hyperref}
\newcommand{\adv}{\mathcal{A}}
\newcommand{\nrnodes}{n}
\newcommand{\nrgroups}{m}
\newcommand{\groups}{\mathcal{G}}
\newcommand{\group}{G}
\newcommand{\outp}{\mathcal{O}}
\newcommand{\gsec}{s}

\newcommand{\secpar}{\sigma}

\makeatletter
\newcommand{\xRightarrow}[2][]{\ext@arrow 0359\Rightarrowfill@{#1}{#2}}
\newcommand{\xLeftarrow}[2][]{\ext@arrow 3095\Leftarrowfill@{#1}{#2}}
\makeatother

\newcommand{\sendright}[1]{$\xrightarrow{\quad #1\quad}$}
\newcommand{\sendleft}[1]{$\xleftarrow{\quad #1\quad}$}

\newcommand{\phase}[1]{
  \multicolumn{3}{l}{\raisebox{-2mm}{\fbox{\textit{#1}}}\hrulefill} \\[3mm]
}

\newcommand{\modp}[1]{#1}

\def\rcs$#1${#1}

\newcommand{\etal}{\textit{et al.}~}	% et al.
\newcommand{\eg}{{e.g.},\ }		% e.g.
\newcommand{\term}[1]{\emph{#1}}		% to define/introduce terms
\newcommand{\cf}{{cf.}\ }               % cf
\newcommand{\ie}{{i.e.},\ }		% i.e.

\newcommand{\assign}{\mathrel{:=}}	% assignment

\title{Private Handshakes\thanks{This research was partially funded by
Sentinels project JASON (NIT.6677).\hfill\penalty-10000
\rcs$Id: secret-handshakes.tex 17 2008-03-31 21:45:57Z jhh $.}}

\author{Jaap-Henk Hoepman} 
\institute{TNO Information and Communication Technology\\
  P.O. Box 1416, 9701 BK \ Groningen, The Netherlands\\
  \email{jaap-henk.hoepman@tno.nl}\\
  and\\
  Institute for Computing and Information Sciences \\
  Radboud University Nijmegen\\
  P.O. Box 9010, 6500 GL \ Nijmegen, the Netherlands\\ 
  \email{jhh@cs.ru.nl}}

\begin{document}

\maketitle

\bibliographystyle{alphacm-}

\begin{abstract}
Private handshaking allows pairs of users to determine which (secret)
groups they are both a member of. Group membership is kept secret to
everybody else. Private handshaking is a more private form of secret 
handshaking~\cite{balfanz2003handshake}, because it does not allow the
group administrator to trace users. 
We extend the original definition of a handshaking protocol to allow
and test for membership of multiple groups simultaneously.
We present simple and efficient protocols for both the single group and
multiple group membership case. 

Private handshaking is a useful tool for mutual authentication, 
demanded by many pervasive applications (including RFID)
for privacy. Our implementations are
efficient enough to support such usually resource constrained scenarios.
\end{abstract}

\section{Introduction}

A secret handshake allows members of a (secret) group to identify each other,
without revealing their membership to potential eavesdroppers or malicious
impostors. As an informal example taken from the real world, it would allow 
FBI agents attending
a hacker convention to recognise each other without giving away their presence
to the rest of the audience\footnote{%
   This, off course, is not withstanding the use of any other distinctive
   features to 'spot' a typical FBI agent. Moreover, in this scenario, where
   all people present belong essentially to just two groups, non-membership of
   one group `proves' membership of the other\ldots
}.

Several years ago, Balfanz \etal~\cite{balfanz2003handshake} revived interest
(\eg~\cite{castellucina2004secrethandshakes}) in the development of secure
(cryptographic) protocols to implement such secret handshakes. According to
them, secret handshakes are fundamentally different from 
\term{one-way accumulators}~\cite{benaloh1993onewayaccumulators} and 
\term{private matchmaking}~\cite{baldwin1985matchmaking,meadows1986matchmaking,zhang-matchmaking}
(not to be confused with distributed match making~\cite{MulV88}).  We show that
this distinction is only superficial (depending on a particular notion of
traitor tracing), and that much simpler protocols, derived
from the literature on matchmaking (and pretty much equivalent to one-way
accumulators) serve equally well as secret handshake protocols.
We call these protocols \emph{private} handshaking protocols.

Such private handshaking protocols (that, unlike secret handshaking, do not
implement traceability)  are quite suitable to resource constrained
environments, like low-end smart card, RFID or NFC-based\footnote{%
	RFID stands for Radio Frequency IDentification. NFC stands for
	Near Field Communication. See the references for more information.
} 
systems~\cite{rankl2003smartcardhandbook,finkenzeller2003rfidhandbook}.
Moreover, they implement a form of mutual authentication that is sorely needed
in many pervasive systems~\cite{weis2003security,hoepman2006crossingborders}. 
For instance, the privacy of a holder of an RFID tag is better protected if the
reader must authenticate to the tag before the tag releases any information. 
A private handshaking protocol could ensure that the tag would only grant
access if the reader and the tag belong to the same group.

\subsection{State of the art}

Private matchmaking protocols, originally studied by Baldwin and
Gramlich~\cite{baldwin1985matchmaking} (and followed up upon by Zhang and
Needham~\cite{zhang-matchmaking}), allow users that share the same `wish' to
locate and identify each other securely and privately.  The canonical example
used in both papers is that of matching job openings at big corporations with
high-ranked managers looking for their next job opportunity. In this example a
corporation will not want to publicly announce availability of a position, and
similarly, a high-ranked manager will not want to reveal his or her job
aspirations to everybody. The protocol of Baldwin and 
Gramlich~\cite{baldwin1985matchmaking} requires the presence of an on-line
trusted third party. Zhang and Needham~\cite{zhang-matchmaking} improve on this
by not using a trusted third party at all, and not using public-key
cryptography either (making their protocol very light-weight).

Secret handshaking protocols, as studied by Balfanz
\etal~\cite{balfanz2003handshake} consider membership of a secret group
instead, and allow members of such groups to reliably identify fellow group
members without giving away their group membership to non-members and
eavesdroppers.  An example of this problem was given in the
introduction. Balfanz \etal{} also pose the additional requirement that a group
member can choose to authenticate to other group members that have a certain
\term{role} within that group.  Furthermore, they require that group membership
is revocable, and that the protocols are forward repudiable, traceable and
collusion resistant (see section \ref{ssec-handshake} for details). Their
protocols are secure under the Bilinear Diffie-Hellman
assumption~\cite{boneh2001idcrypto} 
and
the random oracle model~\cite{bellare1993randomoracles}. 
They require that each user periodically 
obtains fresh pseudonyms from the group administrator, for use in a handshake
protocol run. 

Their results were later improved by
Castelluccia \etal~\cite{castellucina2004secrethandshakes} with protocols based
on CA-Oblivious encryption secure under the random oracle model and either the
Computational Diffie Hellman assumption or the RSA
assumption~\cite{menezes1996handbookcrypto}. Like Balfanz \etal{}, unlinkability in their 
protocols is achieved at the cost of
an ample supply of fresh pseudonyms used one by one in every protocol run.
Also, both protocols assume the existence of a group administrator that 
distributes group secrets to group members, and that can discover any
traitors. Unfortunately, this also implies that the administrator can discover
all instances of a protocol run in which a particular user
participated\footnote{%
  In the current implementations of these protocols, this is trivial because
  the parties exchange pseudonyms initially distributed by the group
  administrator. More fundamentally, this could be achieved in full generality
  by running the traitor tracing protocol on a normal protocol run. By
  definition, this this would reveal the parties involved (provided they were
  members of the group).
}. This is clearly a strong breach of privacy.

Tsudik and Xu~\cite{tsudik2005framework-handshakes} 
extend the secret handshaking problem to more than $2$ participants (but still
determining shared membership of a \emph{single} group), and present protocols
solving this generalisation with reusable credentials. This removes the
main drawback found in previous protocols. Xu and
Yung~\cite{xu2004anonumoushandshakes} previously achieved a similar
reusability of credentials.

%TODO
%- "At crypto wg there was a lot of discussion about the value of traceability;
%so argue this better here!"

Meadows~\cite{meadows1986matchmaking} built a matchmaking protocol without
relying on an on-line trusted third party (but using public key cryptography,
\cf{}~\cite{zhang-matchmaking}). Interestingly, she studied the matchmaking
problem in the secret handshake setting: \ie{} she considered secret group
membership instead of communicating wishes. The difference between both is
subtle, but important (see \cite{balfanz2003handshake}): if the wish can be
guessed, then (by definition of the matchmaking problem that any pair of users
sharing the same wish can identify each other) the owner of that wish can be
identified. Similarly, if `secret' group names are used as input to matchmaking
protocols, then anybody able to guess the group name can locate the other,
real, group members, and moreover can impersonate a group member.

In a similar vain, set intersection
protocols~\cite{freedman2004setintersect,kissner2005setintersect} 
are subtly different
from private handshaking protocols as well. Typically, the domains of the sets
over which the intersection has to be computed is much smaller, and in any
case, any element in the domain is a possible member. For private handshaking
protocols, however, group membership is encoded by a secret value from a much
larger, sparsely occupied, domain. Moreover, not all set intersection protocols
require the outcome of the computation to be secret.
A more thorough discussion of the relationship between secret handshaking,
oblivious encryption/signatures and hidden credentials can be found
in~\cite{holt2005reconciling-handshakes}.

\subsection{Our results}

We define the \term{private handshaking} problem as a more private form of
secret  handshaking~\cite{balfanz2003handshake}, that does not allow a
group administrator (or anyone else) to trace users running the protocol. 
This makes private handshaking a more private form of secret handshaking.
Our model and definitions are described in Sect.~\ref{sec-model}.
The
main contribution of this paper is the conclusion that, when dropping
traceability, much more efficient implementations of handshaking are
possible. This makes such protocols viable for resource constrained
environments, like RFID or NFC-based systems.

We extend the definition of handshake protocols to handle the (much more
common) case where people are members of several groups. Using existing,
single-group, handshaking protocols Alice and Bob (member
of $a$ and $b$ groups respectively) can do no better than running $a \times b$
handshake protocols in parallel to determine all the groups that they share
membership of. We show that, in fact, $O(a+b)$ type protocols exist.

We then present two protocols for private handshaking, one for the case where
Alice and Bob are members of a single group (Sect.~\ref{sec-single}), and
another where Alice and Bob are a member of any number of groups each
(Sect.~\ref{sec-arbitrary}). Both use a single Diffie-Hellman key 
exchange~\cite{DifH76} and exchange as many hashes as the largest number of allowed
group membership per user\footnote{%
  Balfanz \etal~\cite{balfanz2003handshake} argue that a Diffie-Hellman key
  exchange cannot be used to implement secret handshaking. Their argument
  however depends on the requirement that individual members of a group need to
  be traceable, and hence does not apply to \emph{private} handshaking
  protocols. 
}.
Security of the protocols relies on the
Diffie-Hellman assumption~\cite{menezes1996handbookcrypto} and the random oracle 
assumption~\cite{bellare1993randomoracles}.

%TODO
%
%CAN WE PROVE THAT PRIVATE HANDSHAKING IS EQUIVALENT TO SECRET KEY BASED
%AUTHENTICATION?
%(the idea being that any secret key protocol should not leak the secret key,
%hence that key can be used as group secret??)

\section{Model and notation}
\label{sec-model}

\subsection{System and adversary model}

We assume a distributed system of $\nrnodes$ nodes, connected by asynchronous
message passing. Nodes can be members of zero, one or more groups 
$\group \in \groups$. There are $\nrgroups$ different groups.
We write $i \in \group$ if node $i$ belongs to 
group $\group$, and $\groups_i$ for the set of all groups to which node $i$
belongs. We assume group membership is fixed and part of the initialisation of
the system. We will discuss the ramifications of this assumption later on in
Sect.~\ref{sec-concl}.

The system is controlled by a Dolev and Yao~\cite{dolev1981security} style
adversary $\adv$ that may block,
delay, relay, delete, insert or modify messages. This allows
him to force nodes to participate in a protocol run
together with other nodes specified by the adversary\footnote{
  Bellare \etal~\cite{bellare1993authentication,BelPR00} model 
  the same adversarial power by allowing the adversary to query an infinite
  supply of protocol oracles.
}.
The adversary may also corrupt any number of nodes in the system, read 
all data stored by such nodes, and participate in protocol runs 
``being within''
such nodes. Nodes and the adversary are modelled as probabilistic
polynomial-time 
Turing machines. We write $\adv \in \group$ if the adversary corrupted a member
of group $\group$, and $\groups_\adv$ for the set of all groups for which the
adversary corrupted a node. If a node $i$ is corrupted
we write $i \in \adv$. In this case $\groups_i$ is assumed
to be a subset of $\groups_\adv$.
Uncorrupted nodes are honest.

In other words, the adversary induces a sequence of message exchanges and
protocol steps called a \term{run}. At the start of each run, all nodes are
initialised. In this phase, nodes may be given long term secret data needed to
securely run the protocol. However, the adversary may subvert any number of
nodes and retrieve this secret information stored by them. Finally,
the adversary may force any node to reveal any secret information resulting
from a particular protocol exchange. Typically, this involves a session key
established by the protocol.

\subsection{The private handshake problem}
\label{ssec-handshake}

We have the following set of requirements
(\cf\cite{balfanz2003handshake,tsudik2005framework-handshakes}) for a private handshake protocol
run between two nodes $i$ and $j$, belonging to groups  $\groups_i$ and
$\groups_j$ that returns output $\outp_i$ to $i$ and $\outp_j$ to $j$.
All statements below hold with overwhelming probability, 
for arbitrary adversary $\adv$, for an arbitrary
group $\group$ and nodes $i,j$.

\begin{description}
\item[correctness/safety]
  $\outp_i,\outp_j \subseteq \groups_i \cap \groups_j$.
\item[progress]
  If $i$ and $j$ are honest and all messages exchanged between them
  during the run are delivered unaltered, 
  $\outp_i = \outp_j = \groups_i \cap \groups_j$. 
\item[resistance to detection]
  Let $j \in \adv$ but $\adv \notin \group$. Then the adversary $\adv$ cannot
  distinguish a protocol run in which it interacts with
  a node $i \in \group$ from a run involving a simulator\footnote{%
      Note how this requirement subtly circumvents the problem that the
      adversary \emph{does} learn non-membership of $i$ of the groups it is
      itself a member of (by corruption or otherwise).
  }. 
\item[indistinguishability to eavesdroppers]
  Let $i,j \notin \adv$.
  Then the adversary $\adv$
  cannot determine whether $i \in \group$ or $i \notin \group$.
  This holds even if $\adv \in \group$. Note that both participants in the run
  need to be uncorrupted, and that the adversary does not modify\footnote{%
	The powers of the adversary are limited to eavesdropping in this
	case. Clearly, an active adversary belonging to the same group as
	$i$ can stage a man-in-the-middle attack and determine
	membership of $\group$ for $i$ just like a legitimate node $j$
	could.
  }
  messages
  exchanged between $i$ and $j$.
\item[unlinkability]
  Adversary $\adv$ is unable to distinguish a protocol run involving node $i$
  from a protocol run involving a node $j \neq i$ with $\groups_j = \groups_i$,
  even when $\groups_\adv = \groups_i$ and $\adv$ participates in the protocol
  runs\footnote{%
	The statement of this requirement is a bit involved because
	technically, an adversary can distinguish different nodes from the
	groups they are a member of, if the adversary itself is a member of those
	groups and if it participates in the runs. Intuitively, the requirement
	simply says that protocol runs do not carry node identifiers or
	similar. 
  }. 
\item[forward repudiability]
  After the run, node $i$ cannot convince another node $k$ whether 
  $j \in \group$ or not. In other words, a run between 
  $i$ and $j$ is indistinguishable
  from a run between $i$ and $i$, for anyone except $i$.
\end{description}
Traditionally, the following two requirements are listed as well.
\begin{description}
\item[resistance to impersonation]
  Let $j \in \adv$ but $\adv \notin \group$. Then the adversary is not
  able to convince a node $i \in \group$
  that $\adv \in \group$. 
\item[non traceability]
  The group administrator of group $\group$ is unable to link two different
  protocol runs involving the same node $i \in \group$.
\end{description}
However, resistance to impersonation is actually implied by correctness
and the definition of $\groups_i$ when $i$ is corrupted.
And non-traceability is equivalent to unlinkability if the group administrator
is missing (or considered to be a normal, corruptible, node).
We therefore omit these requirements from the list.

We refrain from imposing a fairness requirement
(\cf~\cite{balfanz2003handshake}) which would require
$\outp_i = \outp_j$ always.
Fairness can be guaranteed,
but at the expense of running a complex fair exchange type protocol.

Similarly, we do not require the protocol participants to set up a shared
session key to be used whenever mutual authentication was successful.
The protocols we present, however, do establish such a shared key.

Finally, we note that Meadows~\cite{meadows1986matchmaking} stipulates that
an adversary that has stolen a secret from a group member cannot find out
membership of the someone else without at least revealing group membership.
This is similar to the resistance to impersonation requirement, when fairness
is guaranteed. Otherwise, it will only hold when the adversary initiates the
handshake.
%\cite{zhang-matchmaking}: locating members efficiently

% TODO?
%\subsubsection{Traitor tracing implies loss of privacy}
%
%proof of this

\section{Single membership protocols}
\label{sec-single}

We first present a protocol to determine shared membership of a single group. 
This protocol is basically a Diffie Hellman key exchange using a secret
generator $\gsec$ as the group secret, and using the key validation phase as group
membership test. The validated key can be discarded or used for secure
communication between the authenticated parties. In fact, the protocol is 
very similar to SPEKE~\cite{Jab96}, and Meadows~\cite{meadows1986matchmaking}
basic protocol idea (but without exchanging the secret session key in the clear,
instead using a key verification round as in~\cite{BelPR00}).

\begin{figure}[t]
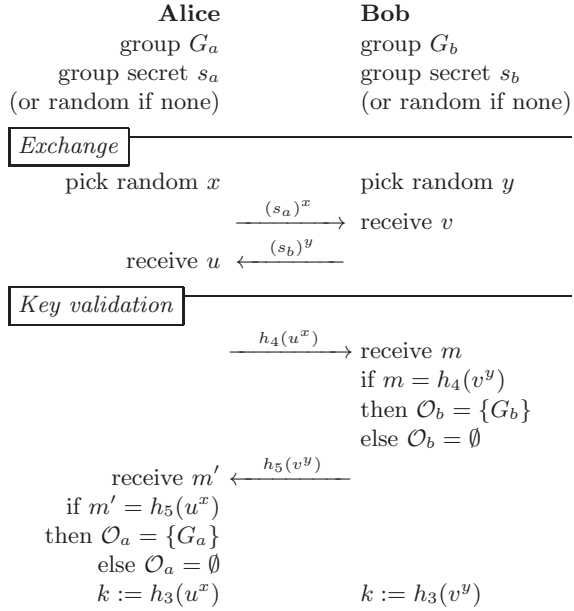

\small
\begin{center}
\begin{tabular}{rcl}
\textbf{Alice} & & \textbf{Bob} \\

group $\group_a$ &			  & group $\group_b$ \\
group secret $\gsec_a$ &                  & group secret $\gsec_b$ \\
(or random if none)&			  & (or random if none)\\

\phase{Exchange}

pick random $x$    &                      & pick random $y$ \\
                   & \sendright{\modp{(\gsec_a)^x}}
                                          & receive $v$\\ 
receive $u$        & \sendleft{\modp{(\gsec_b)^y}}
                                          & \\

\phase{Key validation}

                & \sendright{h_4(\modp{u^x})}
					& receive $m$ \\
		&			& if $m = h_4(\modp{v^y})$ \\
		&			& then $\outp_b = \{ \group_b \}$ \\
		&			& else $\outp_b = \emptyset$ \\
receive $m'$	& \sendleft{h_5(\modp{v^y})} \\
if $m'=h_5(\modp{u^x})$ \\
then $\outp_a = \{ \group_a \}$ \\
else $\outp_a = \emptyset$ \\
$k  \assign h_3(\modp{u^x})$
                &                       & $k \assign h_3(\modp{v^y})$ 
\end{tabular}
\end{center}
\caption{Message flow of the single membership private handshaking protocol.}
\label{fig-shake}
\end{figure}

% TODO: - using match making
% TODO: - using accumulators

\subsection{Security proof}

The following lemmas prove that protocol~\ref{fig-shake} implements
private handshaking. We only sketch the proofs.
Consider an arbitrary run between two nodes $i$ and $j$, belonging to groups
$\groups_i = \{ \group_a \} $ and $\groups_j = \{ \group_b \}$ where 
$i$ returns output $\outp_i$ 
and $j$ returns output $\outp_j$. 
Let  $\adv$ be an arbitrary adversary, and let $\group$ be an arbitrary
group. A property holds with overwhelming probability if it holds with
probability larger than $1-1/2^\secpar$, where $\secpar$ is the security parameter.
It holds with negligible probability if the probability is less than
$1/2^\secpar$.

\begin{lemma}[correctness/safety]
  $\outp_i,\outp_j \subseteq \groups_i \cap \groups_j$
with overwhelming probability.
\end{lemma}

\begin{proof}
Clearly the protocol ensures $\outp_i \subseteq \groups_i$. 
We have $\group_a \in \outp_i$
when $h_5(\modp{u^x}) = h_5(\modp{v^y})$.
This happens only, with overwhelming probability,
when
$\modp{u^x} = \modp{v^y}$, in other words
$\modp{(\gsec_b^y)^x} = \modp{(\gsec_a^x)^y}$.
This holds only with overwhelming probability when
$\gsec_a = \gsec_b$.
\qed
\end{proof}

\begin{lemma}[progress]
  If $i$ and $j$ are honest and all messages exchanged between them
  during the run are delivered unaltered, then
  $\outp_i = \outp_j = \groups_i \cap \groups_j$. 
\end{lemma}

\begin{proof}
This is easily verified by case analysis.
\qed
\end{proof}

\begin{lemma}[resistance to detection]
  Let $j \in \adv$ but $\adv \notin \group$. Then the adversary $\adv$ cannot
  distinguish a protocol run in which it interacts with
  a node $i \in \group$ from a run involving a simulator
  with non-negligible probability.
\end{lemma}

\begin{proof}
% The proof is similar to that of SPEKE~\cite{Jab96}, see e.g.
% MacKenzie~\cite{Mac01b}.
The adversary has to distinguish $\gsec_a^x$ from $g^z$ given $f^x$ for
$f$ known to the adversary, where $x$ is fresh,
random and unknown to the adversary. Moreover, $\gsec_a$ is unknown to the
adversary (but it may know many $\gsec_a^y$, for fresh and unknown $y$, 
from previous protocol runs).
Distinguishing this would violate
the Diffie-Hellman assumption.
\qed
\end{proof}

\begin{lemma}[indistinguishability to eavesdroppers]
  Let $i,j \notin \adv$.
  Then the adversary $\adv$
  cannot determine whether $i \in \group$ or $i \notin \group$ with
  non-negligible probability.
  This holds even if $\adv \in \group$. 
\end{lemma}

\begin{proof}
Similar to the proof of the previous lemma.
\qed
\end{proof}

\begin{lemma}[unlinkability]
  Adversary $\adv$ is unable to distinguish a protocol run involving node $i$
  from a protocol run involving a node $j \neq i$ with $\groups_j = \groups_i$,
  even when $\groups_\adv = \groups_i$ and $\adv$ participates in the protocol
  runs. 
\end{lemma}

\begin{proof}
Nodes $i$ and $j$ share the same state. Hence
all messages sent by $i$ could have been sent by $j$ as well.
\qed
\end{proof}

\begin{lemma}[forward repudiability]
  After the run, node $i$ cannot convince another node $k$ whether 
  $j \in \group$ or not. 
\end{lemma}

\begin{proof}
Because $i$ is a member of $\group$, it can construct a valid protocol run
between $i$ and $j$ all by himself, without $j$ participating at all.
\qed
\end{proof}

\section{Arbitrary membership protocols}
\label{sec-arbitrary}

\begin{figure}[t]
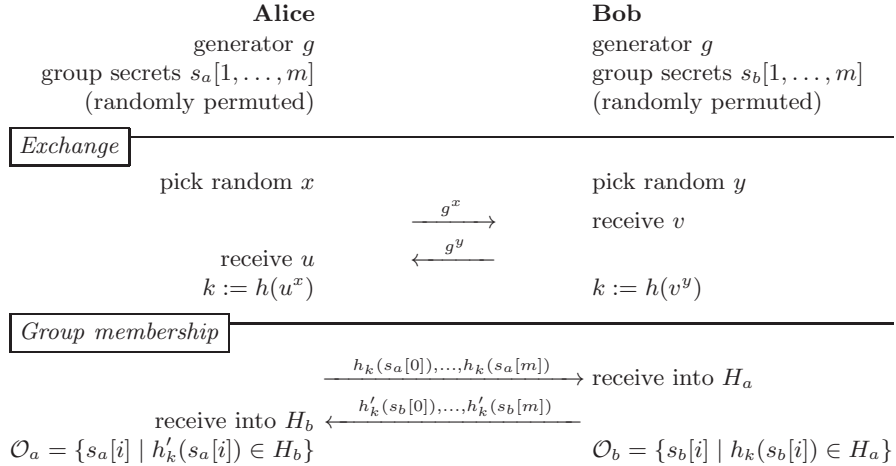

\small
\begin{center}
\begin{tabular}{rcl}
\textbf{Alice} & & \textbf{Bob} \\

generator $g$      &                      & generator $g$ \\
group secrets $s_a[1,\ldots,\nrgroups]$ 
		   &         		  & group secrets $s_b[1,\ldots,\nrgroups]$ \\
(randomly permuted) &			  & (randomly permuted) \\
\phase{Exchange}

pick random $x$    &                      & pick random $y$ \\
                   & \sendright{\modp{g^x}}
                                          & receive $v$\\ 
receive $u$        & \sendleft{\modp{g^y}}
                                          & \\
$k  \assign h(\modp{u^x})$
                &                       & $k \assign h(\modp{v^y})$ \\

\phase{Group membership}

                & \sendright{h_k(s_a[0]),\ldots,h_k(s_a[\nrgroups])}
					& receive into $H_a$ \\
receive into $H_b$
                & \sendleft{h'_k(s_b[0]),\ldots,h'_k(s_b[\nrgroups])} \\

$\outp_a=\{s_a[i] \mid h'_k(s_a[i]) \in H_b\}$
		&			& 
		$\outp_b=\{s_b[i] \mid h_k(s_b[i]) \in H_a\}$ \\
\end{tabular}
\end{center}
\caption{Message flow of the generalised private handshaking protocol.}
\label{fig-arbitrary}
\end{figure}

It is possible to use the single membership protocol to determine
all groups of which both Alice and Bob are a member, by running the previous
protocol for all candidate pairs separately. However, if Alice is a member of
$a$ groups and Bob is a member of $b$ groups, this requires $a \times b$
message exchanges (and more if the number of groups one is a member of should
not be revealed). In this section we describe a more efficient protocol (see
Protocol~\ref{fig-arbitrary}), which does \emph{not} provide traitor
tracing. 

Suppose each user can be a member of at most $\nrgroups$ groups. Each group is
identified by a group secret (which, essentially, is a random value). Each user
$A$ that is a member of a group stores its group secret in an
array $s_A[]$. Any remaining cells in the array are filled with random values
(not corresponding to groups). The array is randomly permuted after
initialisation\footnote{% 
  If not, Bob might be able to infer the number of groups of which Alice is a
  member from the fact that the $x$-th token happens to coincide with a group
  he himself is a member of.
}.
After establishing a shared secret session key $k$ using a Diffie-Hellman key
exchange, Alice and Bob exchange keyed hashes $h_k$ and $h'_k$ of each group
secret. Real implementations should use HMAC~\cite{bellare1996hmac}.
Alice stores the hashes it receives in a set $H_B$, looks for entries
in $s_A[]$ whose hash occurs in $H_B$, and adds those as common group members
to $G_A$.

Note that Alice needs to use a hash function different from the one used by
Bob, in order to avoid detection of shared membership by eavesdroppers.
If Alice wishes not to reveal membership of certain groups, she can replace the
corresponding group secret with a random value. However, Bob cannot avoid
revealing his membership of those groups (unless he decides to do so
independently from Alice).

\subsection{Security proof}

The following lemmas prove that protocol~\ref{fig-arbitrary} implements
private handshaking for multiple group. We sketch the proofs of the lemmas.
Consider an arbitrary run between two nodes $i$ and $j$, belonging to groups
$\groups_i$ and $\groups_j$ where 
$i$ returns output $\outp_i$ 
and $j$ returns output $\outp_j$ (where we treat the group secrets
$s_i[x]$ to represent their respective groups). 
Let  $\adv$ be an arbitrary adversary, and let $\group$ be an arbitrary
group. A property holds with overwhelming probability if it holds with
probability larger than $1-1/2^\secpar$, where $\secpar$ is the security parameter.
It holds with negligible probability if the probability is less than
$1/2^\secpar$.

\begin{lemma}[correctness/safety]
  $\outp_i,\outp_j \subseteq \groups_i \cap \groups_j$
with overwhelming probability.
\end{lemma}

\begin{proof}
Clearly $\outp_i \subseteq \groups_i$. If $x \in \outp_i$ then also
$h'_k[x] \in H_j$. Hence
$h'_k(x)=z$ for some $z$ received in the second phase of the protocol.
If $z$ is not sent by $j$, then $k$ is unknown to the adversary. Hence
the chances that $h'_k(x)=z$ are negligible.
If $z$ is sent by $j$ then $z=h'_k(s_j[y])$ for some $y$. This happens with
overwhelming probability if $x=s_j[y]$ and hence $x \in \groups_j$.
\qed
\end{proof}

\begin{lemma}[progress]
  If $i$ and $j$ are honest and all messages exchanged between them
  during the run are delivered unaltered, then
  $\outp_i = \outp_j = \groups_i \cap \groups_j$. 
\end{lemma}

\begin{proof}
This is easily verified by case analysis.
\qed
\end{proof}

\begin{lemma}[resistance to detection]
  Let $j \in \adv$ but $\adv \notin \group$. Then the adversary $\adv$ cannot
  distinguish a protocol run in which it interacts with
  a node $i \in \group$ from a run involving a simulator
  with non-negligible probability.
\end{lemma}

\begin{proof}
Since $j \in \adv$, the adversary does know the shared session key
$k$ derived using the Diffie-Hellman key exchange. However, since
$\adv \notin \group$, it does not know the secret $s_i[x]$ for group $\group$.
Hence it cannot tell whether $h_k(s_i[x])$ and $h_{k'}(s_i[x])$ are hashes
for the same group exchanged during different sessions, or if these hashes
correspond to different groups. This holds even if the adversary knows
$k'$ for the other session as well.
\qed
\end{proof}

\begin{lemma}[indistinguishability to eavesdroppers]
  Let $i,j \notin \adv$.
  Then the adversary $\adv$
  cannot determine whether $i \in \group$ or $i \notin \group$ with
  non-negligible probability.
  This holds even if $\adv \in \group$. 
\end{lemma}

\begin{proof}
If $i,j \notin \adv$, then the adversary does not know the shared session key
$k$ derived using the Diffie-Hellman key exchange. With a fresh, unknown,
random key $k$, the keyed hash value $h_k(s_i[x])$ corresponding to
the secret for group $\group$ is indistinguishable from a random value, even
if the adversary knows $s_i[x]$.
\qed
\end{proof}

\begin{lemma}[unlinkability]
  Adversary $\adv$ is unable to distinguish a protocol run involving node $i$
  from a protocol run involving a node $j \neq i$ with $\groups_j = \groups_i$,
  even when $\groups_\adv = \groups_i$ and $\adv$ participates in the protocol
  runs. 
\end{lemma}

\begin{proof}
Nodes $i$ and $j$ share the same state. Hence
all messages sent by $i$ could have been sent by $j$ as well.
\qed
\end{proof}

\begin{lemma}[forward repudiability]
  After the run, node $i$ cannot convince another node $k$ whether 
  $j \in \group$ or not. 
\end{lemma}

\begin{proof}
Because $i$ is a member of $\group$, it can construct a valid protocol run
between $i$ and $j$ all by himself, without $j$ participating at all.
\qed
\end{proof}

\section{Conclusions}
\label{sec-concl}

We have presented two efficient protocols for secret handshaking. The second
protocol efficiently supports membership of more than one group.
The focus in this work is the efficiency of the protocols. They use only a few,
quite simple, operations. This may allow the implementation of these protocols
on resource constrained devices, like perhaps higher-end RFID tags.
It is especially in these kinds of environments that a form of mutual
authentication is required to provide a certain level of security and/or
privacy. 

Our protocols do not allow for easy revocation of group membership: all
remaining members need to be given a new, fresh, group secret. More efficient
ways to support group membership revocation are an interesting topic for
further research, especially given the requirement that the resulting protocols
should still be efficient and should not allow a group adminstrator to trace
users. We also wish to develop more formal proofs for the
security of our protocols.

Two other possible extensions of the basic pairwise private handshake are left
for further investigation. First of all, one could consider
a private group handshake where a subgroup of a secret group can recognise
membership of the same group simultaneously (\eg when setting up a meeting).
Secondly, one could create password based private handshakes by using the 
original idea of Jablon~\cite{Jab96}
based on a passkey shared by the members of the group.

We thank Flavio D. Garcia, David Galindo and Berry Schoenmakers for fruitful
discussions on this topic, and the anonymous referees for their very insightful
comments and suggestions. 

\bibliography{/home/jhh/work/lit/bib/strings,secret-handshakes}

\end{document}